\documentclass[12pt]{article}

\usepackage{epsf,rotate} 
\usepackage{subeqn} 
\usepackage{cite}

\newcommand{\gev}{\,\mathrm{GeV}}
\newcommand{\errorpm}[2]{
\raisebox{-0.5ex}{\shortstack[l]{$\scriptstyle+#1$\\$\scriptstyle-#2$}}}
\newcommand{\as}{\alpha_s}
\newcommand{\eps}{\varepsilon}
\newcommand{\kkm}{\ensuremath{\mathrm{K^0\!-\!\ov{K^0}}\,}-mixing}
\newcommand{\bbm}{\ensuremath{\mathrm{B^0\!-\!\ov{B^0}}\,}-mixing}
\newcommand{\kkmd}{\ensuremath{\mathrm{K_L\!-\!K_S}\,}-mass difference}
\newcommand{\ov}{\overline}
\newcommand{\dstwo}{\ensuremath{\mathrm{|\Delta S| \!=\!2}}}
\newcommand{\dsone}{\ensuremath{\mathrm{|\Delta S| \!=\!1}}}
\newcommand{\lt}{\left}
\newcommand{\rt}{\right}
\newcommand{\text}[1]{\mathrm{#1}}
\newcommand{\eq}[1]{(\ref{#1})}
\newcommand{\pr}{Phys.\ Rev.\ }
\newcommand{\prd}{\pr D}
\newcommand{\np}{Nucl.\ Phys.\ }
\newcommand{\npb}{\np B}
\newcommand{\prl}{\pr Lett.\ }
\newcommand{\pl}{Phys.\ Lett.\ }

\newcommand{\fig}[1]{Fig.~\ref{#1}}
\newcommand{\oll}{\ensuremath{Q_{S2}}}
\newcommand{\laMSb}{\ensuremath{\Lambda_{\overline{\mathrm{MS}}}}}
\newlength{\miniwidth}
\newlength{\miniwidthplot}
\newlength{\nseparation}
\newenvironment{nfigure}[1]
        {\begin{figure}[#1]\hrule\vspace{\nseparation}\par}  
        {\vspace{\nseparation}\par \hrule \end{figure}}     

\begin{document}
\renewcommand{\baselinestretch}{1.25}
TUM-HEP-254/96 \hfill hep-ph/9609310

~\vspace{2.43truecm}
\begin{center}
{\Large PHENOMENOLOGY OF $\mathbf{\eps_K}$ IN THE TOP
ERA}\footnote{\parbox[t]{15truecm}{Invited Talk at the workshop on K physics, 
Orsay, France, 30th May -- 4th June 1996.\\[-0.4\baselineskip]
Work supported by BMBF under contract no.~06-TM-743.}}\\[2\baselineskip]
\textsl{Ulrich
Nierste\footnote{e-mail:Ulrich.Nierste@feynman.t30.physik.tu-muenchen.de},
Physik-Department, TU M\"unchen, D-85747 Garching, Germany}
\end{center}
\vfill
\renewcommand{\baselinestretch}{1}
\begin{center}
\textbf{\large Abstract} 
\end{center}
Todays key information on the shape of the unitarity triangle is
obtained from the well-measured quantity $\eps_K$ characterizing the
CP-violation in \dstwo\ transitions.  The phenomenological analysis
requires the input of four key quantities: The magnitudes of the CKM
elements $V_{cb}$ and $V_{ub}$, the top quark mass and the
non-perturbative parameter $B_K$. In the recent years all of them 
have been determined with increasing precision.  In order to keep up
with this progress the \dstwo-hamiltonian had to be obtained in the
next-to-leading order (NLO) of renormalization group improved
perturbation theory.  I present the NLO results for the QCD
coefficients $\eta_1$ and $\eta_3$, which have been calculated by
Stefan Herrlich and myself, and briefly sketch some aspects of
the calculation.  Then I give an update of the unitarity triangle
using the summer 1996 data for the input parameters. The results for
the improved Wolfenstein parameters $\ov{\rho}$ and $\ov{\eta}$ and
the CKM phase $\delta$ are
\begin{eqnarray}
-0.21 \leq \ov{\rho} \leq 0.22, \qquad 
0.27\leq \ov{\eta} \leq 0.43, && \qquad 
57^\circ \leq \delta \leq 122^\circ .   
\nonumber
\end{eqnarray}
The range for the quantity $\sin ( 2 \beta)$ entering CP asymmetries
in B-decays is found as 
\begin{eqnarray}
0.46 \; \leq & \sin \lt( 2 \beta \rt) & \leq \; 0.79 .
\nonumber 
\end{eqnarray}
The given ranges correspond to one standard deviation in the 
input parameters. Finally I 
briefly discuss the \kkmd.

\newpage

\section{Motivation}
\renewcommand{\baselinestretch}{1.25}\normalsize $\eps_K$
characterizes the CP-violation in the mixing of the neutral Kaon
states $\mathrm{K^0}$ and $\mathrm{\ov{K^0}}$.  This indirect
CP-violation has been discovered in 1964 by Christenson, Cronin, Fitch
and Turlay \cite{ccft}. In the subsequent three decades refined
experiments have reduced the error in $\eps_K$ below 1\%
\cite{ad,pdg}, but yet no other CP-violating quantity has been
unambiguously determined. In the Standard Model the only source of
CP-violation is a complex phase $\delta$ in the
Cabibbo-Kobayashi-Maskawa (CKM) matrix. Hence today the measured value
of $\eps_K$ plays the pivotal r\^{o}le in the determination of
$\delta$. In the near future B-physics experiments will reveal whether
the single parameter $\delta$ can simultaneously fit CP-violating
observables in both the B- and the K- system and will eventually open
the door to new physics.
 
The lowest order contribution to the $|\Delta S|\!=\!2$-amplitude
inducing \kkm\ is depicted in \fig{box}.  
\begin{nfigure}{tb}
\centerline{\epsfxsize=6cm \epsffile{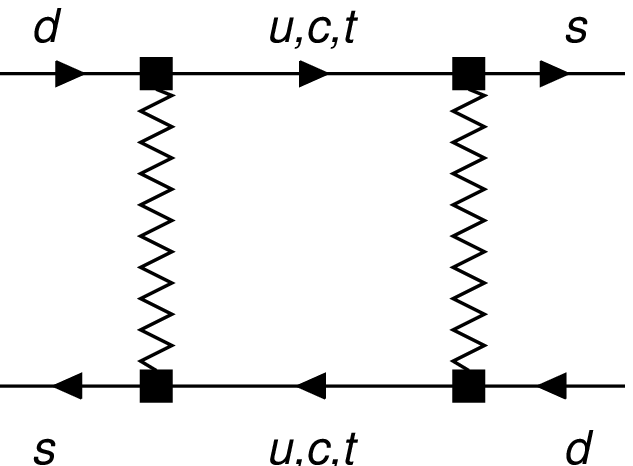}}
\caption[$\Delta S\!=\!2$ box diagram]{
\renewcommand{\baselinestretch}{0.95}\normalsize\slshape
        The lowest order box diagram mediating a \dstwo\ transition.
        The zigzag lines denote W-bosons or
        fictitious Higgs particles.}
\label{box} 
\end{nfigure}
In order to calculate the Standard Model prediction for $\eps_K$ one
must first separate the short distance physics from long distance
effects in the \dstwo\ transition amplitude.  After successively
integrating out the heavy degrees of freedom $m_t$, $M_W$ and $m_c$
one ends up with an effective low-energy \dstwo-hamiltonian:
\begin{eqnarray}
H^{|\Delta S|=2} &=&
               \frac{ G_{F}^2 }{ 16 \pi^2 } M_W^2  \left[
                  \lambda_c^2 \eta_1^{\star} x_c^{\star} 
                   \! + \!
                  \lambda_t^2 \eta_2^{\star}  
                  S ( x_t^{\star} ) 
                 \! + \!  
                2 \lambda_c \lambda_t \eta_3^{\star} 
                  S(x_c^{\star} , x_t^{\star}  )
                   \right]  
 b(\mu) Q_{S2}(\mu) + \text{h.c.} \; \; \label{s2}
\end{eqnarray} 
Here $G_{F}$ is the Fermi constant, $M_W$ is the W boson mass,
$\lambda_j=V_{jd} V_{js}^{*}$ comprises the CKM-factors and $Q_{S2} $
is the local \dstwo\ four-quark operator
\begin{eqnarray}
\oll &=& 
 \overline{s}_j \gamma_\mu (1-\gamma_5) d_j \cdot 
 \overline{s}_k \gamma^\mu (1-\gamma_5) d_k  \label{ollintro} 
\end{eqnarray}
with $j$ and $k$ being colour indices.  $ x_q^{\star}
=m_q^{\star\,2}/M_W^2$, where $m_q^{\star}=m_q (m_q) $, $q=c,t$, are
running quark masses in the $\overline{\text{MS}}$ scheme.  The
Inami-Lim functions $S(x)$ and $S(x,y)$ contain the quark mass
dependence of the box diagram in \fig{box}.

The short distance QCD corrections are comprised in the coefficients
$\eta_1$, $\eta_2$ and $\eta_3$ with a common factor $b(\mu)$ split
off.  They are functions of the charm and top quark masses and of the
QCD scale parameter $\Lambda_{\text{QCD}}$.  Further they depend on
the \emph{definition} of the quark masses used in the Inami-Lim
functions: In (\ref{s2}) the $\eta_i$'s are defined with respect to
$\overline{\text{MS}}$ masses $m_q^{\star}$ and are therefore marked
with a star. In the absence of strong interaction one has $\eta_i
b(\mu)=1$.

$|\eps_K|$ is proportional to the imaginary part of the hadronic
matrix element $\langle \ov{K^0} \mid H^{\dstwo} \mid {K^0} \rangle$.
It thereby involves the hadronic matrix element of $\oll$ in 
\eq{ollintro}, which is conveniently parametrized as
\begin{eqnarray}
\langle   \ov{K^0} \mid \oll (\mu) \mid {K^0} \rangle &=&
\frac{8}{3} \frac{B_K}{ b (\mu )} f_K^2 m_K^2 .    
\label{bk}
\end{eqnarray}

Here $m_K$ and $f_K$ are mass and decay constant of the neutral K
meson and $\mu$ is the renormalization scale at which the short
distance calculation of \eq{s2} is matched with the non-perturbative
evaluation of \eq{bk}.  $B_K$ in \eq{bk} is defined in a
renormalization group (RG) invariant way, because the $\mu$-dependent
terms from \eq{bk} and \eq{s2} cancel in
$\langle \ov{K^0} \mid H^{\dstwo} \mid {K^0} \rangle$.

Now the CKM matrix depends on four independent parameters. The
convenient Wolfenstein parametrization expands all CKM elements
in terms of the well-known quantity 
$\lambda \simeq 0.22\simeq |V_{us}|$ to order $\lambda^3$. 
The proper study of CP violation, however, requires a higher accuracy
\cite{sch,blo,hn3}. The improved Wolfenstein approximation adopted 
in \cite{blo,hn3} yields 
\begin{eqnarray}
V_{us}=\lambda+O \lt( \lambda^7 \rt), && \quad 
V_{cb}= A \lambda^2+O \lt( \lambda^8 \rt), \qquad
V_{ub}=  A \lambda^3 \lt( \rho - i \eta  \rt). \nonumber
\end{eqnarray}
From $b\rightarrow c$ decays one extracts $|V_{cb}|$ and thereby $A$.
The information encoded in the remaining two parameters $(\rho,\eta)$
is traditionally depicted as a \emph{unitarity triangle} in the
complex plane. Two of its corners are located at $(0,0)$ and $(1,0)$,
while the exact location $(\ov{\rho},\ov{\eta})$ of its top corner is
defined by
\begin{eqnarray}
\ov{\rho} \,+\, i \, \ov{\eta} &=& 
-\frac{V_{ud} V_{ub}^*}{V_{cd} V_{cb}^*} .
\label{rhobar}
\end{eqnarray}
$\ov{\rho}$ and $\ov{\eta}$ are related to $\rho$ and $\eta$ by
\cite{blo,hn3}
\begin{eqnarray}
\ov{\rho} = \rho \, 
                 \left( 1- \frac{\lambda ^2}{2} + O (\lambda^4)
                 \right) ,
&\qquad&
\ov{\eta} = \eta \, 
            \left( 1- \frac{\lambda ^2}{2} + O (\lambda^4) \right) .
\nonumber
\end{eqnarray}
Inserting the improved Wolfenstein approximation into the expression 
for $|\eps_K|$ yields
\begin{eqnarray}
&& \hspace{-15mm}
5.3 \cdot 10^{-4} = B_K A^2 \ov{\eta} 
    \left[ \lt( 1 - \ov{\rho} + \Delta  \lt(\ov{\rho},\ov{\eta}
             \rt) \rt) 
         A^2 \lambda^4 \eta^{\star}_2 S(x^{\star}_t)
           + \eta^{\star}_3 S(x^{\star}_c,x^{\star}_t)       
           - \eta^{\star}_1 x^{\star}_c  \right] 
        \lt( 1+ O \lt(\lambda^4\rt) \rt) .
         \label{cons2}  
\end{eqnarray}
In the absence of the small term 
\begin{eqnarray}
\Delta  \lt( \ov{\rho},\ov{\eta} \rt) &=& \lambda^2 \, \lt(
        \ov{\rho} -\ov{\rho}^2 -\ov{\eta}^2 \rt) 
\end{eqnarray}
\eq{cons2} defines a hyperbola in the $\lt( \ov{\rho},\ov{\eta}
\rt)$-plane.  The second input needed to determine the
shape of the unitarity triangle is provided by the measured value of
$|V_{ub}/V_{cb}|$, which fixes a circle in the
$(\ov{\rho},\ov{\eta})$-plane: 
\begin{eqnarray}
\left| \frac{V_{ub}}{V_{cb}} \right| &=& 
      \lambda \sqrt{\ov{\rho}^2+\ov{\eta}^2}
             \,  \lt( 1 + \frac{\lambda^2}{2} + 
                    O\lt( \lambda^4 \rt)  \rt). \label{circ}
\end{eqnarray}
The intersection points of circle and hyperbola are the allowed values
for $(\ov{\rho},\ov{\eta})$ (see \fig{fig:hyperbola}).
\begin{nfigure}{tb}
\centerline{\epsfxsize=10cm \epsffile{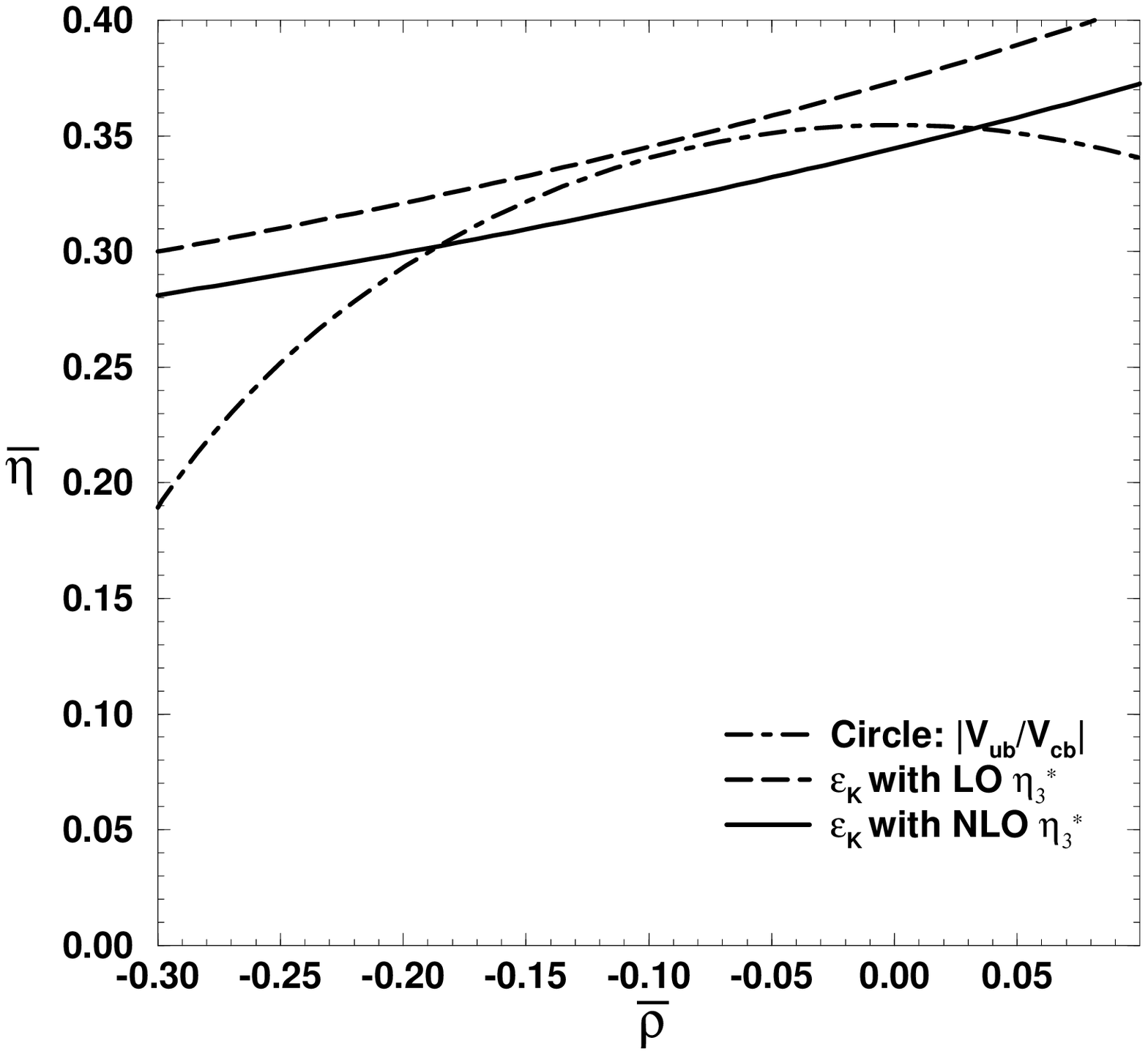}}
\vspace{-5mm}
\caption[]{ \renewcommand{\baselinestretch}{0.95}\normalsize
\slshape
The two solutions for $(\bar{\rho},\bar{\eta})$ are the intersections
of the hyperbola from $\eps_K$ with the circle obtained from 
$|V_{ub}|/|V_{cb}|$. The plot also visualizes the impact of the NLO 
calculation of $\eta_3$ on the unitarity triangle: For the 
chosen set of input parameters the LO value for $\eta_3$ yields 
no solution for $(\bar{\rho},\bar{\eta})$.}\label{fig:hyperbola}
\end{nfigure}

The standard phenomenological analysis of $\eps_K$ involves four
key input parameters: The hyperbola \eq{cons2} is entered by $B_K$, $m_t$
and (via $A$) $|V_{cb}|$ and the circle involves $|V_{ub}/V_{cb}|$.
In the past few years significant progress has been made in the
determination of these quantities:
\begin{itemize}
\item $|V_{cb}|$: Both exclusive and inclusive $b \rightarrow c \ell
      \nu_\ell$ decays have been precisely measured by CLEO and
      ALEPH. The theoretical extraction of $|V_{cb}|$ from the
      decay rates has been refined by the development of heavy quark
      effective theory and today we know $|V_{cb}|$ to 8 \% accuracy.
\item $|V_{ub}/V_{cb}|$: In addition to inclusive measurements of
      $b\rightarrow u\ell \nu_\ell$ decays now also the exclusive
      decays $B \rightarrow \rho \ell \nu_\ell$ and $B \rightarrow \pi
      \ell \nu_\ell$ have been measured by CLEO.
\item $B_K$: The lattice results have steadily improved as reported 
      by G.~Kilcup at this  workshop.
\item $m_t$: Most importantly the top quark has been discovered at FERMILAB. 
      In the time before the top discovery the 
      unknown value of $m_t$ was the largest source of
      uncertainty in the phenomenology of $\eps_K$. Now in the 
      top era the experimental error in $m_t$ affects 
      the determination of the unitarity triangle less than those
      in the other three input parameters.
\end{itemize} 
Clearly the accuracy of the QCD coefficients $\eta_1$,
$\eta_2$ and $\eta_3$ of $H^{|\Delta S|=2}$ entering
the hyperbola \eq{cons2} must keep up with this progress! This  
has required to calculate them in the \emph{next-to-leading order}\/ (NLO)
of renormalization group (RG) improved perturbation theory. 

\section{$\mathbf{H^{|\Delta S|=2}}$ in the next-to-leading order}
Suppose one would try to determine the strength of \dstwo\ transitions
simply by calculating the box diagram of \fig{box} and its dressing
with gluons. This would result in a very poor description of \kkm\ for
various reasons: First long-distance QCD is non-perturbative and
therefore cannot be described by the exchange of gluons. Second the
true external states are mesons rather than quarks. Third the largely
separated mass scales in the problem induce large logarithms in the
radiative corrections: For example the QCD corrections to the box
diagram with two internal charm quarks contain $(\alpha_s/\pi) \ln x_c
\simeq 1$, which spoils perturbation theory. Fourth one faces a scale
ambiguity for the same reason: Should one evaluate the running
coupling $\alpha_s$ at the scale $\mu=M_W$, $\mu=m_c$ or any scale in
between? These problems can be overcome with the help of Wilson's
operator product expansion, which expresses the Standard Model
amplitudes of interest in terms of an effective hamiltonian, in which
the fields describing heavy particles such as W-boson and top-quark do not
appear anymore. Instead the transitions mediated by them are described
by effective operators, which can be obtained from the Standard Model
diagrams by contracting the heavy lines to a point. E.g.\ contracting
the top-quark and the W-boson lines in \fig{box} yields the
four-quark operator $\oll$ in \eq{ollintro}. The operators are
multiplied by short distance Wilson coefficients, which are
functions of the heavy masses. Disturbingly large logarithms can then
be summed to all orders by applying the RG to the
coefficients. Starting with the heaviest masses $m_t$ and $M_W$, the
whole procedure is then repeated with the next lighter particle, in
our case the charm quark. Finally in $H^{|\Delta S|=2}$ the 
coefficient of the operator $\oll$ is rewritten in terms of the 
Inami-Lim functions and the $\eta_i$'s.

The minimal way to incorporate short distance QCD effects is the
leading logarithmic approximation. For example the leading order (LO)
expression for $\eta_1\cdot x_c$ contains the result of the box
diagram in \fig{box} with two internal charm quarks and the large
logarithmic term $[\alpha_s \ln x_c ]^n$ to all orders
$n=0,1,2,\ldots$ of the perturbation series.  The NLO improves the LO
results by including those terms with an additional factor of
$\alpha_s$, in the case of $\eta_1$ these are the results of the
two-loop diagrams with an additional gluon dressing the box and the
summation of $\alpha_s [\alpha_s \ln x_c ]^n$, $n=0,1,2,\ldots$.  In
the modern formalism described in the first paragraph the LO
$\eta_i$'s have been calculated by Gilman and Wise in 1983 \cite{gw}
partly confirming earlier results obtained with different methods. Yet
in general LO results suffer from various conceptual drawbacks. In the
case of the $\eta_i$'s one faces four problems:
\begin{itemize}
\item The LO results do not reproduce the correct dependence on $m_t$.
      Especially the dependence of $\eta_3$ on $m_t$ enters in the
      NLO.
\item Likewise the proper definition of $m_t$ is a NLO
      issue. One must go to the NLO to learn how to use the FERMILAB
      measurement of the pole quark mass $m_t^{\text{pole}}$ in a low
      energy hamiltonian like $H^{|\Delta S|=2}$ in \eq{s2}. In a NLO
      expression it is appropriate to use the one-loop formula to
      relate $m_t^{\text{pole}}$ to $m_t^{\star}$ entering $H^{|\Delta
      S|=2}$. These two definitions $m_t^{\text{pole}}$  and 
      $m_t^{\star}$ differ by 8 GeV, which is more than the 
      present experimental error in $m_t^{\text{pole}}$. 
\item The fundamental QCD scale parameter $\laMSb$ is an essential NLO 
      quantity and cannot be used in LO expressions.
\item The LO results for $\eta_1$ and $\eta_3$ suffer from large 
      renormalization scale uncertainties. Their reduction requires 
      a  NLO calculation. 
\end{itemize}
The coefficient $\eta_2$ has been calculated in the NLO by Buras, Jamin
and Weisz \cite{bjw}. The NLO order results for $\eta_1$ \cite{hn1}
and $\eta_3$ \cite{hn3,hn4} have been derived by Stefan Herrlich and
myself. 

The three results read
\begin{eqnarray}
\eta_1^{\star} &=& 1.31\errorpm{0.25}{0.22}\,,
\quad \quad 
\eta_2^{\star} \; = \; 0.57\errorpm{0.01}{0.01}\,,
\quad \quad 
\eta_3^{\star} \; = \;  0.47\errorpm{0.03}{0.04}\,.
\label{coneta}
\end{eqnarray} 
The coefficients are scheme independent except that they depend on the
definition of the quark masses in $H^\dstwo$. The errors are estimated
from the remaining scale uncertainty. $\eta_2^\star$ and
$\eta_3^\star$ depend on $\as$ and the quark masses only marginally.
The quoted value for $\eta_1^\star$ corresponds to $m_c^\star=1.3 \gev$
and $\as (M_Z) =0.117 \gev$. For other values of $\as$ and $m_c^\star$
see the tables in \cite{hn1,hn4}.

The NLO values in \eq{coneta} have to be compared with the old LO
results:
\begin{eqnarray}
\eta_1^{\mathrm{LO} } & \approx & 0.74 \, ,
\quad \quad
\eta_2^{\mathrm{LO} } \approx  0.59 \, , 
\quad \quad
\eta_3^{\mathrm{LO} } \approx  0.37 \, .
\label{old}
\end{eqnarray} 
If one takes the difference between \eq{old} and \eq{coneta} as an
estimate of the inaccuracy of the LO expressions, one finds that the
use of the $\eta_i^{\mathrm{LO} }$'s in \eq{cons2} imposes an error
onto the phenomenological analysis of $\eps_K$ which is comparable in
size to the error stemming from the hadronic uncertainty in $B_K$.
 
I close the theoretical part of my talk by briefly sketching some
details of the calculation of $\eta_1$ and $\eta_3$: The new feature 
compared to other NLO calculations was the appearance of two-loop
diagrams with the insertion of \emph{two} \dsone\
operators. Pictorially they are obtained by contracting the W-lines
in \fig{box} to a point and dressing the diagram with gluons. 
Here the proper renormalization of such Green's functions with two
operator insertions had to be worked out \cite{hn4}. This has required the 
correct renormalization of so called \textsl{evanescent operators},
which appear in the context of dimensional regularization \cite{hn2}.  
Such operators induce a new type of scheme dependence into the
calculation, which of course cancels in physical observables 
\cite{hn2,hn4}. 

\section{1996 phenomenology of $\mathbf{\eps_K}$ and the 
         $\mathbf{K_L\!-\!K_S}\,$-mass difference}\label{sect:ph}
The first phenomenological analysis of $\eps_K$ with NLO precision has
been presented in \cite{hn3}. In \cite{hn3} $\ov{\rho}$ and
$\ov{\eta}$ defined in \eq{rhobar}, the CKM phase $\delta$ and other
quantities related to the CKM matrix are tabulated as a function of
$B_K$, $m_t$, $|V_{cb}|$ and $|V_{ub}/V_{cb}|$. Here I will update the
unitarity triangle with the actual values of these key input
parameters.
\begin{nfigure}{tb}
\centerline{\epsfxsize=14cm \epsffile{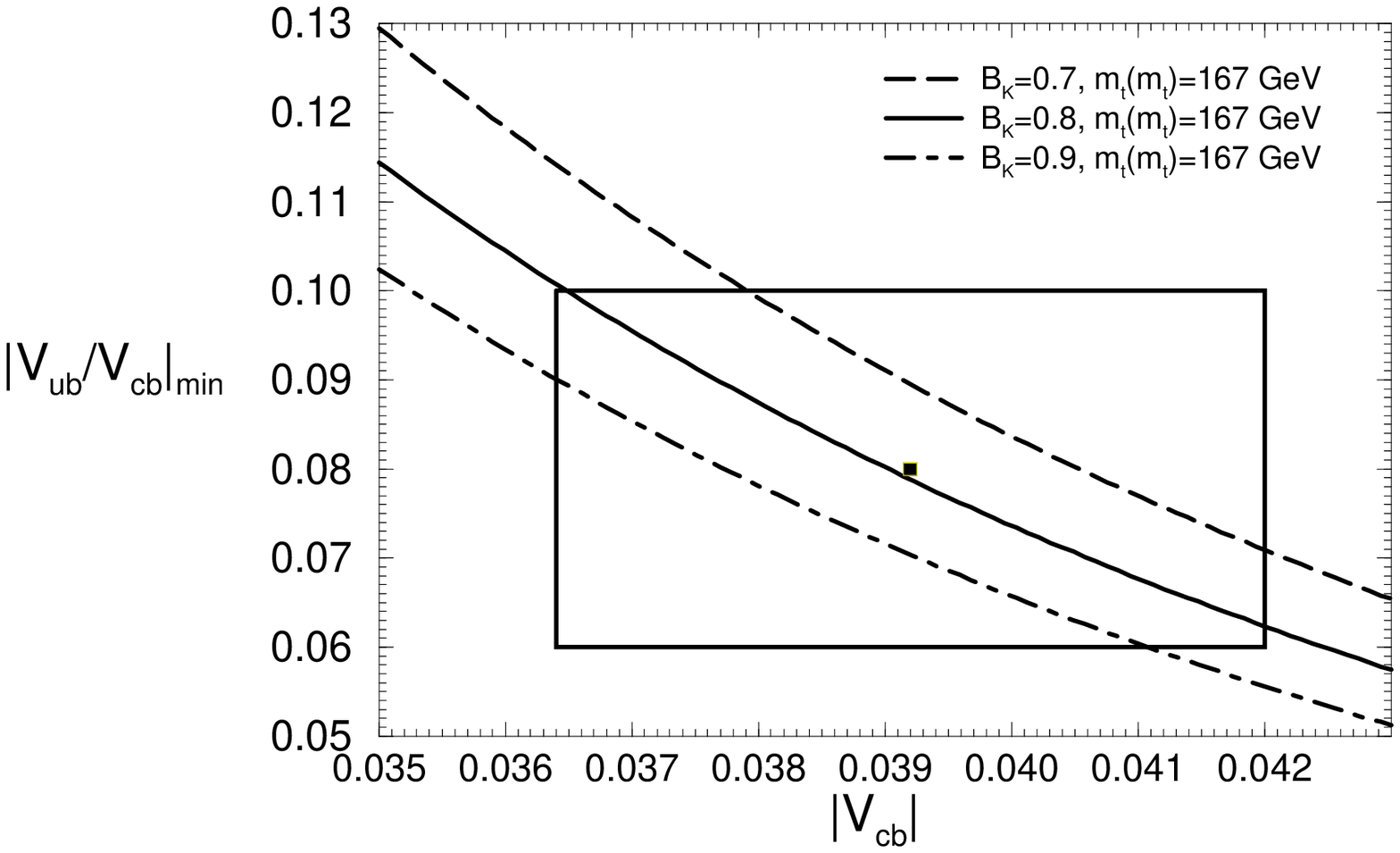}}
\caption[]{\renewcommand{\baselinestretch}{0.95}\normalsize
 \slshape 
Borderline of new physics: For each pair $(m_t^{\star},B_K)$ the
measured value for $\varepsilon_K$ defines a curve. The points below
the curve are excluded, if $\eps_K$ is solely due to Standard Model
physics.  The rectangle shows the limits in \eq{pars} for $|V_{cb}|$ and
$|V_{ub}/V_{cb}|$ obtained from tree-level b-decays. The central
values used in the analysis are marked with the small filled square.
          }\label{fig:lobo}
\end{nfigure}

The existence of a solution for  $(\ov{\rho},\ov{\eta})$ requires that
the hyperbola in \eq{cons2} intersects or at least touches the circle
defined in \eq{circ} as shown in \fig{fig:hyperbola}. This feature 
yields lower bounds on each of the four input parameters as a function 
of the other three ones. In \fig{fig:lobo} this condition is displayed
as a constraint on the CKM elements. The present status of the
unitarity triangle is shown in \fig{fig:ut}. 
\begin{nfigure}
\centerline{\epsfxsize=13cm \epsffile{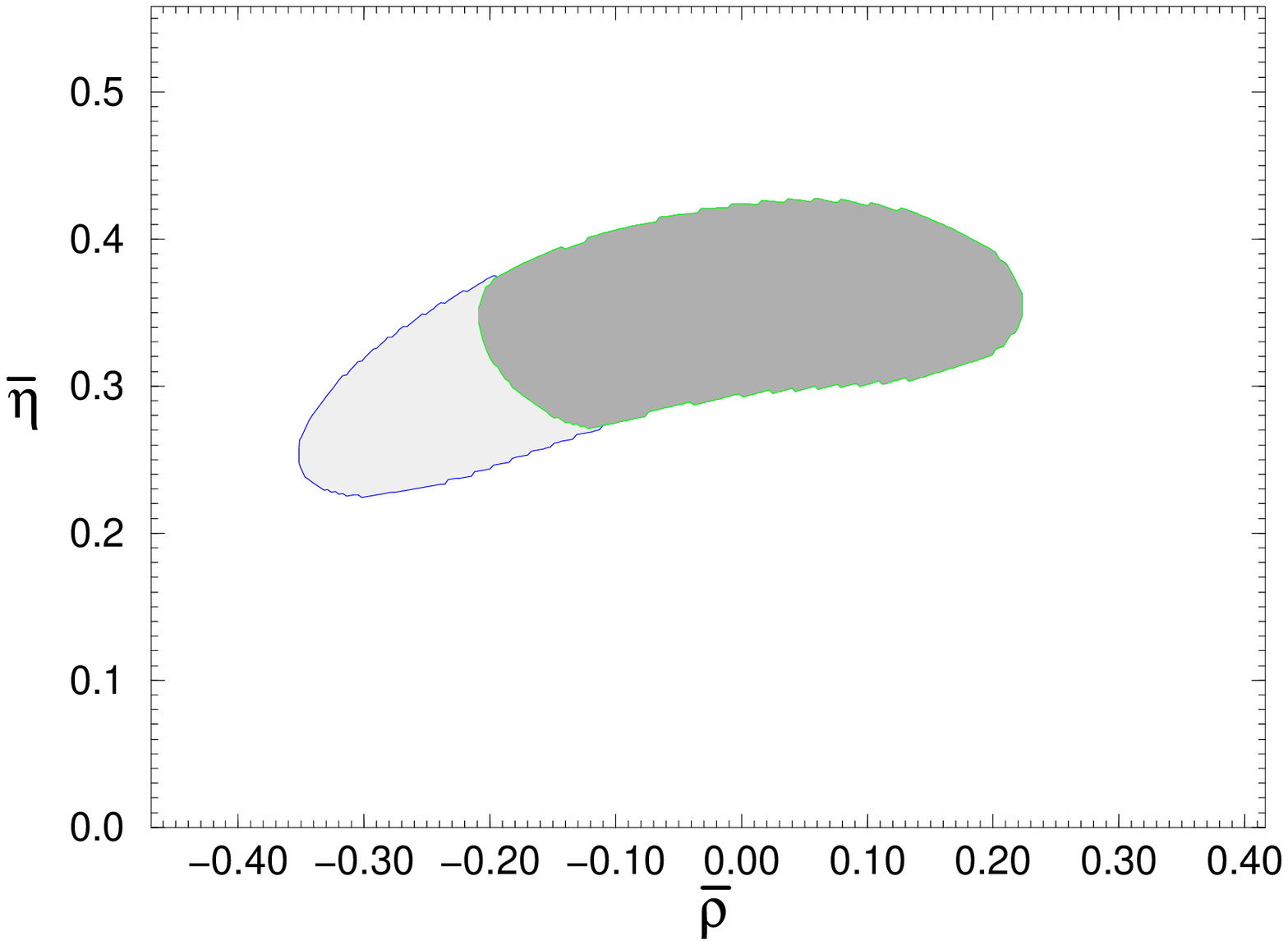}}
\caption[]{ \renewcommand{\baselinestretch}{0.95}\normalsize 
\slshape  The allowed region for the top $(\ov{\rho},\ov{\eta})$ of 
          the unitarity triangle: The dark shade denotes the area 
          which is simultaneously allowed by $\eps_K$ and 
          ${ B^0\!-\!\ov{B^0}}\,$-mixing.
          The light gray area complies with $\eps_K$, but not
          with $\Delta m_{B_d}$. The impact of the limit on 
          $\Delta m_{B_s}$ is discussed in the text.
          }\label{fig:ut}
\end{nfigure}
The input parameters are taken as \cite{wa}
\begin{eqnarray}
\hspace{-10mm} &&  |V_{cb}| = 0.0392\pm0.0028 , \quad 
\left| \frac{V_{ub}}{V_{cb}}\right| = 0.08\pm0.02, \quad 
m_t^{\star} = (167 \pm 6 ) \gev, \quad B_K = 0.8 \pm 0.1 .
\label{pars}
\end{eqnarray}
Here $|V_{cb}|$ is extracted from an analysis of exclusive
semileptonic B-decays. The quoted value for $m_t^\star$ corresponds to
$m_t^{\text{pole}}=(175\pm 6 )\gev$. The range for $B_K$ includes the
ballpark of the lattice results presented in \cite{wa,lk} and the result
of the $1/N_c$ expansion in \cite{bbg}.

Next we include the experimental information from \bbm\ into our 
analysis: The ALEPH results \cite{wa} 
\begin{eqnarray}
\Delta m_{B_d} = (0.464\pm0.018)\, \text{ps}^{-1} = (305\pm 12)
\,\mu\text{eV}, &&\qquad \Delta m_{B_s} > 9.2 \, \text{ps}^{-1},
\label{pars2}
\end{eqnarray}
exclude a part of the region allowed by $\eps_K$. The measured value 
of $\Delta m_{B_d}$ constrains the distance of $(\ov{\rho},\ov{\eta})$
to the point $(1,0)$:
\begin{eqnarray}
(1-\ov{\rho})^2 + \ov{\eta}^2 &=& 
   \frac{4.76 \cdot 10^8 \gev \, \cdot \, \Delta m_{B_d}}{B_{B_d} \,
      F_{B_d}^2 \,  S(x_t^\star ) \, |V_{cb}|^2 } . \label{dmd}
\end{eqnarray}
In \fig{fig:ut} we have used \cite{wa,acc}
\begin{eqnarray}
F_{B_d} = (175 \pm 30) \text{MeV}, \quad && \qquad  B_{B_d} = 1.31 
\label{pars3}
\end{eqnarray}
and the central value for $\Delta m_{B_d} $ in \eq{pars2}. The
variation of 30 MeV in $F_{B_d}$ accounts for the actual error of 25
MeV and the smaller errors in $B_{B_d}$ and $\Delta m_{B_d} $ reported
in \cite{wa}. The whole shaded area in \fig{fig:ut} shows the
region which is allowed from the analysis of $\eps_K$ alone. The
analysis of $\eps_K$ is not particularly sensitive to the treatment of
the errors in \eq{pars}. In \fig{fig:ut} they have been treated
statistically: Setting
$(x_1,x_2,x_3,x_4)=(|V_{cb}|,|V_{ub}/V_{cb}|,m_t^\star,B_K)$ and
denoting their central values by $\ov{x}_i$ and their errors by
$\Delta x_i$ the $x_i$'s have been restricted to the
$1\sigma$-ellipsoid $\sum_i (x_i-\ov{x}_i)^2/(\Delta x_i)^2\leq 1$.
The error in the analysis of $\Delta m_{B_d}$ is theoretical and
therefore treated non-statistically: For each point
$(\ov{\rho},\ov{\eta})$ it has been checked whether it corresponds to
a value of $F_{B_d}$ in the range given in \eq{pars3}. Finally the
bound for $\Delta m_{B_s}$ also excludes a part of the light gray
area, but it does not further constrain the dark region allowed from
both $\eps_K$ and $\Delta m_{B_d}$, if one uses
$F_{B_s}/F_{B_d}=1.25\pm 0.10$, which one expects from an unquenched
lattice calculation \cite{wa}. Yet future tighter bounds on $\Delta
m_{B_s}$ will give extra information on the unitarity triangle
\cite{bbl}.  From the result for $(\ov{\rho},\ov{\eta})$ one can
extract the CKM phase $\delta$.  Another interesting quantity is $\sin
( 2 \beta)$, where $\beta$ is the angle of the unitarity triangle
adjacent to the corner $(1,0)$. $\sin ( 2 \beta)$ enters CP
asymmetries in B-decays. One finds
\begin{eqnarray}
57^\circ \leq \delta \leq 122^\circ , && \qquad  
0.46 \leq \sin \lt( 2 \beta \rt) \leq 0.79 .
\end{eqnarray}

The coefficient $\eta_1$ is known less accurately than $\eta_2$ and
$\eta_3$ due to the sizeable scale uncertainty in \eq{coneta} and its
strong dependence on $\as$. Fortunately the term involving
$\eta_1$ in \eq{cons2} is of minor importance for the analysis of
$\eps_K$. In contrast the short distance part of the \kkmd, which is
obtained from the real part of $\langle \ov{K^0} \mid H^{\dstwo} \mid
{K^0} \rangle$, is dominated by $\eta_1$ and therefore plagued by
theoretical uncertainties. With $B_K$ in \eq{pars} and
$\as(M_Z)=0.118\pm0.004$ the ratio of the short distance part 
of $\Delta m_K$ and its experimental result \cite{pdg,ad2} reads
\begin{eqnarray}
\frac{\left(\Delta m_K\right)_{\text{SD}}}
        {\left(\Delta m_K\right)_{\text{exp}}}
&=& 0.74 \errorpm{0.25}{0.20} . \label{sd}
\end{eqnarray}
At least this reveals a short distance dominance of $\Delta m_K$ in 
accordance with the expectations from power counting \cite{hn1,hn3}. 

\subsubsection*{Acknowledgements}
I am grateful to the organizers for the invitation and a supporting
grant allowing me to attend this excellent workshop.  During this
conference I have enjoyed many stimulating discussions, especially
with Stefano Bertolini, Eduardo de Rafael, Jean-Marc~G\'erard, Fred
Gilman and Guido Martinelli.  Section~\ref{sect:ph} has benefited from
discussions with Andrzej Buras, Stefan Herrlich and Weonjong Lee.


\begin{thebibliography}{1xx}\renewcommand{\baselinestretch}{0.95}\normalsize 
\bibitem{ccft}  J.\ H.\ Christenson, J.\ W.\ Cronin, 
        V.\ L.\ Fitch and R.\ Turlay,
        \prl 13 (1964) 138.\\
        J.\ H.\ Christenson, J.\ W.\ Cronin, 
        V.\ L.\ Fitch and R.\ Turlay, 
        \pr 140B (1965) 74.
\bibitem{ad} CPLEAR collaboration, R.~Adler \emph{et al.}, \pl B363
(1995) 243. 
\bibitem{pdg} Particle Data Group, \pr D54 (1996) 1.
\bibitem{sch} M.~Schmidtler and K.~Schubert, Z.~Phys.~C53 (1992) 347. 
\bibitem{blo} A.~J.~Buras, M.~E.~Lautenbacher and G.~Ostermaier, \pr D50
              (1994) 3433. 
\bibitem{hn3} S.~Herrlich und U.~Nierste, \prd 52 (1995) 6505.
\bibitem{gw} F.\ J.\ Gilman and M.\ B.\ Wise, \prd 27 (1983) 1128.
\bibitem{bjw} A.\ J.\ Buras, M.\ Jamin and P.\ H.\ Weisz, \npb 347 (1990) 491.
\bibitem{hn1} S.~Herrlich und U.~Nierste, \npb 419 (1994) 292.
\bibitem{hn4} S.~Herrlich und U.~Nierste, \textsl{The Complete 
      $|\Delta S| \!=\!2$-Hamiltonian in the Next-To-Leading Order}, 
      \npb, \textsl{in press}, hep-ph/9604330.
\bibitem{hn2} S.~Herrlich und U.~Nierste,
        \npb 455 (1995) 39.
\bibitem{wa} Talks by P.~Tipton, J.~Flynn and L.~Gibbons  at
         the \textsl{28th Int.~conference on high energy physics, Warsaw, 
                     Poland, July 1996}.  
\bibitem{lk} W.~Lee and M.~Klomfass, hep-lat/9608089.
\bibitem{bbg} W.~A.~Bardeen, A.~J.~Buras and J.-M.~G\'erard, 
    \pl\/  B211 (1988) 343. \\ 
    J.-M.~G\'erard, Acta Phys.~Pol.~B21 (1990) 257.
\bibitem{acc} C.~R.~Allton, M.~Ciuchini, M.~Crisafulli, V.~Lubicz and 
              G.~Martinelli,\\ \np B431 (1994) 667.
\bibitem{bbl} G.~Buchalla, A.~J.~Buras and M.~E.~Lautenbacher, 
             hep-ph/9512380. \\  
            A.~Ali and D.~London, hep-ph/9607392.
\bibitem{ad2} CPLEAR collaboration, R.~Adler \emph{et al.}, \pl B363
(1995) 237. 

\end{thebibliography}
\end{document}